\documentclass{vgtc}                          % final (conference style)
\ifpdf%                                % if we use pdflatex
  \pdfoutput=1\relax                   % create PDFs from pdfLaTeX
  \usepackage{graphicx}                % allow us to embed graphics files
  \DeclareGraphicsExtensions{.pdf,.png,.jpg,.jpeg} % for pdflatex we expect .pdf, .png, or .jpg files
\else%                                 % else we use pure latex
  \usepackage{graphicx}                % allow us to embed graphics files
  \DeclareGraphicsExtensions{.eps}     % for pure latex we expect eps files
\fi%

\graphicspath{{figures/}{pictures/}{images/}{./}} % where to search for the images

\usepackage{microtype}                 % use micro-typography (slightly more compact, better to read)
\PassOptionsToPackage{warn}{textcomp}  % to address font issues with \textrightarrow
\usepackage{textcomp}                  % use better special symbols
\usepackage{mathptmx}                  % use matching math font
\usepackage{times}                     % we use Times as the main font
\usepackage{cite}                      % needed to automatically sort the references
\usepackage{tabu}                      % only used for the table example
\usepackage{booktabs}                  % only used for the table example
\onlineid{1181}

\vgtccategory{Research}

\usepackage{orcidlink}

\vgtcinsertpkg

\title{Expanding Targets in Virtual Reality Environments: A Fitts' Law Study}

\author{Rongkai Shi~\orcidlink{0000-0001-8845-6034}\thanks{e-mail: rongkai.shi19@student.xjtlu.edu.cn} %
\and Yushi Wei~\orcidlink{0000-0002-6003-0557}\thanks{e-mail: yushi.wei21@student.xjtlu.edu.cn} %
\and Yue Li~\orcidlink{0000-0003-3728-218X}\thanks{e-mail: yue.li@xjtlu.edu.cn}
\and Lingyun Yu~\orcidlink{0000-0002-3152-2587}\thanks{e-mail: lingyun.yu@xjtlu.edu.cn}
\and Hai-Ning Liang~\orcidlink{0000-0003-3600-8955}\thanks{Corresponding author; e-mail: haining.liang@xjtlu.edu.cn}
}
\affiliation{\scriptsize Department of Computing, School of Advanced Technology \\ Xi'an Jiaotong-Liverpool University}

\abstract{Target pointing selection is a fundamental task. According to Fitts' law, users need more time to select targets with smaller sizes. Expanding the target to a larger size is a practical approach that can facilitate pointing selection. It has been well-examined and -deployed in 2D user interfaces. However, limited research has investigated target expansion methods using an immersive virtual reality (VR) head-mounted display (HMD). In this work, we aimed to fill this gap by conducting a user study using ISO 9241-411 multi-directional pointing task to examine the effect of target expansion on target selection performance in VR HMD. Based on our results, we found that compared to not expanding the target, expanding the target width by 1.5 and 2.5 times during the movement can significantly reduce the selection time. We hope that the design and results derived from the study can help frame future work. 
} 

\CCScatlist{
  \CCScatTwelve{Human-centered computing}{Human computer interaction (HCI)}{Interaction paradigms}{Virtual reality};
  \CCScatTwelve{Human-centered computing}{Human computer interaction (HCI)}{Empirical studies in HCI}{}
}

\begin{document}

\maketitle
\section{Introduction}
%% \section{Introduction} %for journal use above \firstsection{..} instead
Target pointing selection is a fundamental interaction task in all user interfaces (UIs). Expanding the target to a larger size is a practical approach that can facilitate pointing selection. In desktops, tablets, and other 2D UIs, users commonly need to point to the interface widgets, such as buttons and icons. Expanding the size of the target can improve the target selection performance. McGuffin and Balakrishnan~\cite{McGuffin2002Acquisition, McGuffin2005Fitts} showed empirical evidence that target expansion during users reaching the cursor to the target could reduce the pointing time. Except for the linear expansion, researchers also developed non-linear magnification effects, such as fisheye views~\cite{Gutwin2002Improving,Hornbaek2007Untangling} or Voronoi-based expansions~\cite{Guillon2014Static,Guillon2015Investigating}, to support the target selection in 2D UIs. 

While in 3D UIs, users are no longer pointing to 2D interface elements but to 3D objects. However, limited research has investigated target expansion methods in 3D UIs. Early in 2007, Vanacken~\cite{Vanacken2007Exploring} presented 3D bubble cursor to support target selection in dense 3D environments. Once the 3D bubble cursor captured a spherical target, a second, expanding sphere around the captured target would appear to be contained by the cursor. Argelaguet and Andujar~\cite{Argelaguet2008Improving} proposed two target expansion strategies for 3D selection tasks in CAVE. Both research studies focused on expanding targets to support selection in dense and occluded virtual environments. 

With recent advancements, virtual reality (VR) head-mounted displays (HMDs) have become powerful and popular. VR HMDs offer users an immersive experience that is largely greater than other devices. The interaction in VR HMDs also differs from other devices. Users normally control the ray emitted from the 6-DOF controller (i.e., ray-casting~\cite{Bowman1997Evaluation}) to point to the target objects in virtual environments provided by VR HMDs. Yu et al.~\cite{Yu2018Target} tested Expanding Target technique in VR HMDs and found that the selection performance was improved by expanding the size of the pointing object. However, to the best of our knowledge, no research has been conducted to investigate the effect of target expansion in Fitts' law tasks in VR HMDs. This can help understand and predict user behavior to aid designers and developers of VR environments in providing target expansion techniques to enable efficient and accurate target selection in different VR application scenarios.

Fitts' law is one of the most famous HCI predictive models for human movements~\cite{Fitts1954Information} and has recently been studied in immersive VR environments (e.g.,~\cite{Batmaz2022Improving}). Fitts' law states that the movement time to acquire a target is affected by the distance to and the size of the target. One of the most widely used mathematical variations is defined in ISO 9241-411:2012~\cite{ISO}: 

\begin{equation} \label{Eq:Fitts}
Movement Time = a+b*log_2(\frac{A}{W}+1)=a+b*ID
\end{equation}

where $A$ is the effective target distance, $W$ is the effective target size, and the $log$ term represents the effective index of difficulty ($ID$).\footnote{Effective target distance, size, and index of difficulty are usually expressed as $A_e$, $W_e$, and $ID_e$. To improve readability, we directly use $A$, $W$, and $ID$ to represent effective measures in this paper.} The coefficients $a$ and $b$ are derived from linear regression using the empirical results. According to ISO standard~\cite{ISO}, $ID$ and movement time can be further used to calculate the throughput: 

\begin{equation} \label{Eq:TP}
Throughput = (\frac{ID}{Movement Time})
\end{equation}

In this work, we adapted the multi-directional pointing task described in the ISO standard~\cite{ISO} to investigate the effect of target expansion on target selection performance in VR environments. We used movement time, error rate, and throughput as measurements to examine the selection performance. Besides, we also recorded the selection endpoints and visualized their distributions. We conducted a user study ($N=17$) to compare three expansion factors (expanding the target width by 1.5, 2, or 2.5 times during the movement) against a baseline condition (no expansion) in the Fitts' law task. Our results showed that expanding expansion factors of 1.5 and 2.5 can significantly reduce the movement time. This work mainly made an empirical contribution based on a user study with seventeen participants. We hope future work on target expansion and Fitts' law in VR environments can be framed based on this work's exploration.

\begin{figure}[tb]
 \centering 
 \includegraphics[width=\columnwidth]{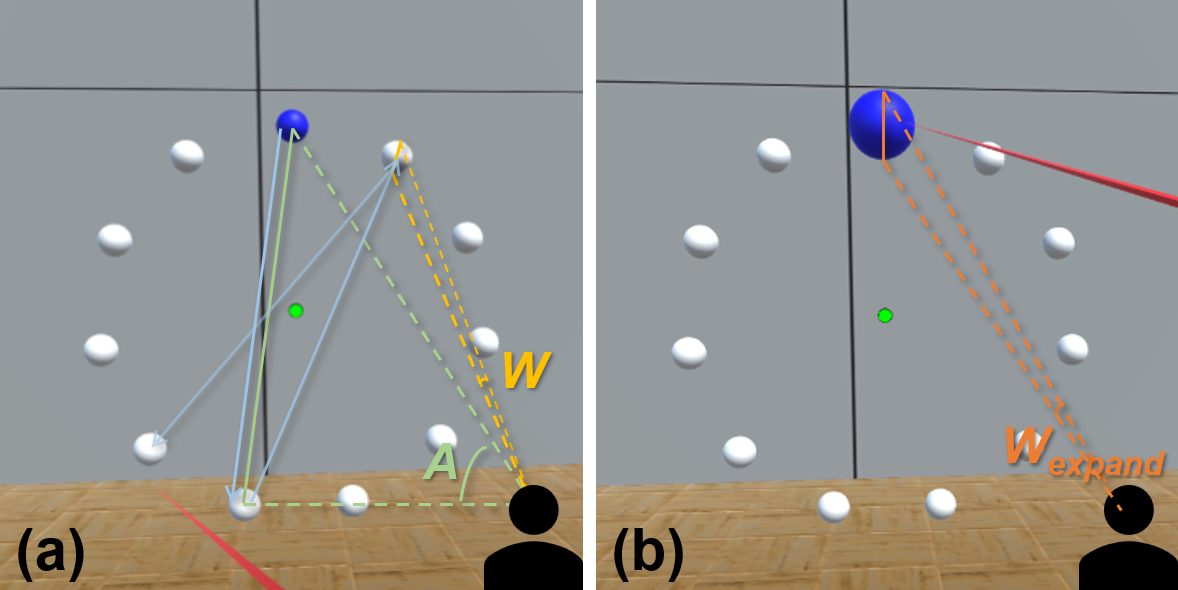}
 \caption{Experimental setup. (a) $W$ is the target width and $A$ is the movement amplitude. The blue arrows indicate the trial order. (b) $W_{expand}$ represents the expanded target width. $W$, $A$, and $W_{expand}$ are all effective distances using angular forms. The green button in the center represents the start button. }
 \label{fig:exp}
\end{figure}

\section{Preliminary Study}
We aimed to investigate the effect of target expansion on target selection. To do so, we conducted a user study with seventeen participants recruited from a local university and used a Meta Quest 2 VR HMD to provide the testing environment. Our primary research question was \textit{how expanding targets would affect the speed, accuracy, and throughput of target selection}. 

% \subsection{Participants and Apparatus}
% Seventeen participants %(XX females, XX males; aged between XX to XX; $M$ = XX, $SD$ = XX) 
% from a local university volunteered to participate in this study. Based on the results of a pre-experiment questionnaire, all of them had at least some VR experience before. They were all right-handed and had a normal or corrected-to-normal vision. 

% We used a Meta Quest 2 VR HMD to provide the testing environment. Meta Quest 2 has a 1832$\times$1920 per-eye resolution, an 89° horizontal field of view, and a 120 Hz refresh rate. Two Quest 2 controllers were used for the controls in the testing environment. 

\subsection{Task and Stimuli}
We used an ISO 9241-411:2012~\cite{ISO} multi-directional tapping task with 11 grey spherical targets placed as a circle (Fitts' ring design). Participants were asked to use the dominant-hand controller to point at the targets and press the trigger on the non-dominant-hand controller to perform the selection. This setting was made to avoid the Heisenberg effect~\cite{bowman2001using}. They were informed to select the target as fast and as precisely as possible. The task started with the target at the top of the ring after participants triggered the start button located at the center of the ring. Once participants selected a target, the next target was on the diagonal, and continued clockwise on the ring. The goal target would be in blue as an indication. Participants were sitting in the middle of an empty virtual room where visual depth cues were provided via this setting. 

The original target width was $W$, and the movement amplitude was $A$. When participants controlled the ray-casting to move past a specified fraction $P$, that is, the expansion point, of the total distance $A$, the target size $W$ would instantaneously expand to $W_{expand}$. In this study, we set $P=0.9*A$~\cite{McGuffin2005Fitts}. \autoref{fig:exp} shows the experimental setup. Following the approach described by Yu et al.~\cite{Yu2019Modeling}, we used a spherical coordinate system and angular representations whenever feasible. Thus, $W$, $A$, and $W_{expand}$ were all effective distances.

\subsection{Experimental Design}
We used three target widths $W$ (1.5°, 2.5°, and 3.5°) and two movement amplitudes $A$ (15° and 20°), which led to six indices of difficulty (\textit{IDs}). This study followed a 6$\times$4 within-subject design with two factors: index of difficulty $ID$ (2.402, 2.747, 2.807, 3.170, 3.459, 3.841) and expansion factor $E$ (1, 1.5, 2, and 2.5). An expansion factor of 1 represented a baseline condition where the target size would not change during the trial. While for the remaining three conditions, the target size would expand---the target width would increase to 1.5, 2, and 2.5 times the initial width. The orders of $W$ and $E$ were counterbalanced using a Latin Square approach separately. For a given $W$, the order of $A$ was randomized. The data from the first trial for each $ID$ condition (i.e., each $W \times A$ combination or each ring) was discarded. Thus, in total, we recorded 6 indices of difficulty $\times$ 4 expansion factors $\times$ 10 repetitions $\times$ 17 participants = 4080 trials of data.

\subsection{Measurements}
For each trial, we recorded the movement time (second, s) and endpoint $(x, y)$. The endpoints were also represented using angular forms~\cite{Yu2019Modeling}. We further calculated error rates (\%) and throughput (°/s) for each condition per participant. Throughput was calculated using \autoref{Eq:TP}.

% A spherical coordinate system $(r, \theta, \phi)$ was used to define a point. Similar to~\cite{Yu2019Modeling}, we ignored the distance $r$ because the ray-casting can be regarded as infinite. Thus, $p = (x, y)$ can be used to present an endpoint, where $x$ is the angular error distance parallel to the direction of movement (from the starting target to the goal target), $y$ is the angular error distance perpendicular to the line of movement (as shown in Figure). The origin of the two axes was set at the center of the goal target. 

% The depth of the Fitts' ring was always the same; that is, the distance between participants and the targets in all trials was a constant value. \textcolor{red}{[Depth=100m? from Difeng's paper.]} 

% \begin{table}[tb]
%   \caption{Randomized task variables for each condition (W = target width, A = movement amplitude, ID = index of difficulty, and E = expansion factor).}
%   \label{tab:ID}
%   % \scriptsize
% 	\centering%
%   \begin{tabu}{%
% 	r%
% 	*{6}{c}%
% 	*{1}{r}%
% 	}
%   \toprule
%      &  & \multicolumn{4}{c}{\textit{ID}}\\
%     \textit{W} & \textit{A} & $E$=1 & $E$=1.5 & $E$=2 & $E$=2.5\\
%     \midrule
%     1.5 & 15 & 3.459 & 2.939 & 2.585 & 2.322\\
%     2.5 & 15 & 2.807 & 2.322 & 2 & 1.766\\
%     3.5 & 15 & 2.402 & 1.948 & 1.652 & 1.441\\
%     1.5 & 20 & 3.841 & 3.306 & 2.939 & 2.663\\
%     2.5 & 20 & 3.17  & 2.663 & 2.322 & 2.07\\
%     3.5 & 20 & 2.747 & 2.266 & 1.948 & 1.716\\
%   \bottomrule
%   \end{tabu}%
% \end{table}

\subsection{Procedure}
Before the experiment, participants were briefly introduced to the experimental setup, including the device, task, and procedure. After that, they started a 3-minute training session to familiarize themselves with the VR device and ISO pointing task. Then they began to complete the formal trials. The whole experiment lasted about 35 minutes per participant.

% \section{}
% \subsection{}
% \subsubsection{}
% \paragraph{}

\subsection{Results and Discussion}

\subsubsection{Movement Time, Error Rate, and Throughput}
\begin{figure}[tb]
 \centering 
 \includegraphics[width=\columnwidth]{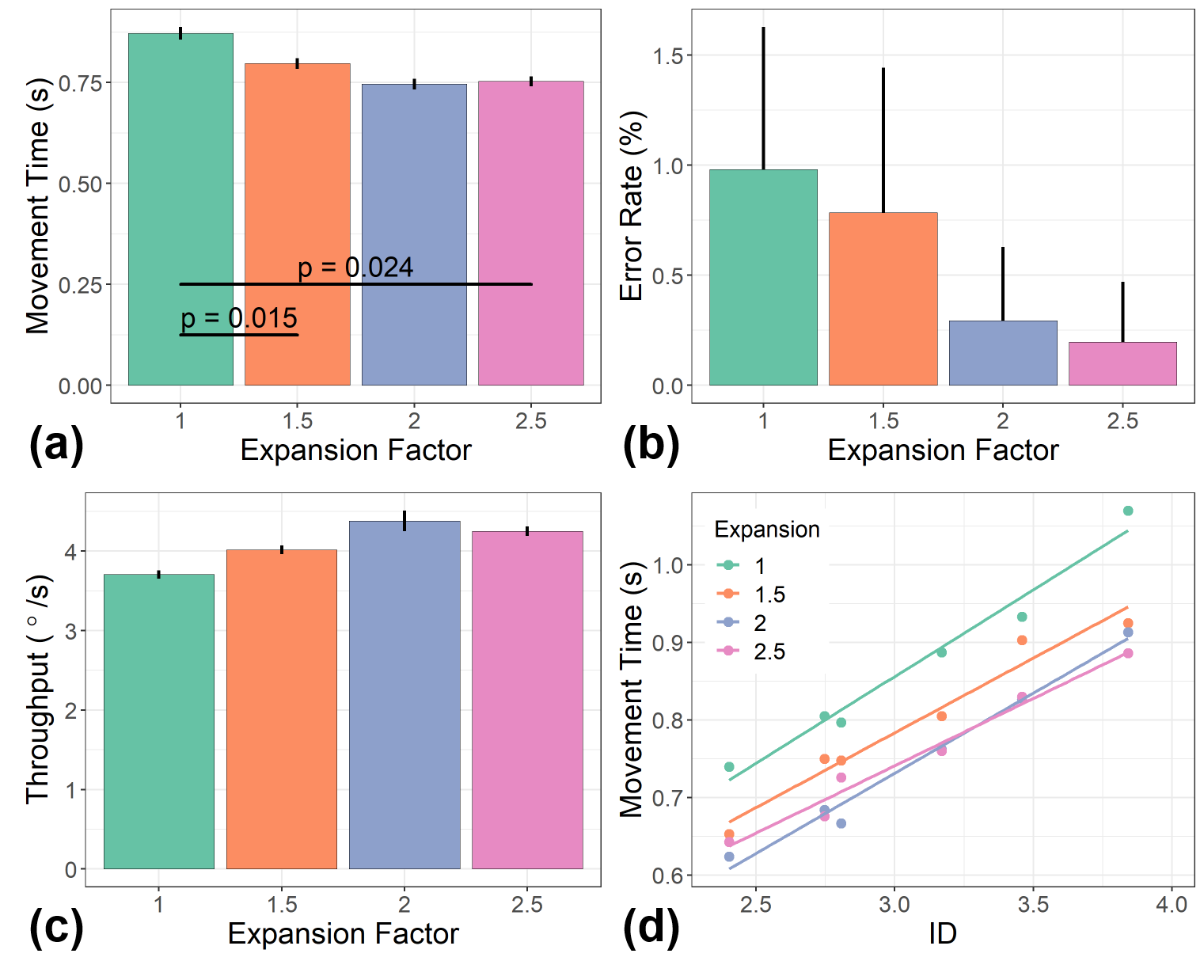}
 \caption{(a-c) Bar plots for Movement Time (a), Error Rate (b), and Throughput (c). Error bars in (a-c) represent 95\% confidence intervals. (d) Fitts' law results for the four expansion factors. }
 \label{fig:barplotsandlm}
\end{figure}

We used repeated measure (RM-) ANOVA to analyze the movement time, error rate, and throughput results. We applied Greenhouse-Geisser correction for degrees of freedom whenever needed ($\epsilon>.75$ for our cases) and Bonferroni correction for post hoc analyses. Results are summarized in \autoref{fig:barplotsandlm}(a-c) and \autoref{tab:RM-ANOVA}. 

We only found a significant main effect of \textit{ID} ($F(2.466,39.460)=3.776$, $p=.024$, $\eta_p^2=.191$) and a significant main effect of \textit{E} on movement time ($F(3,48)=6.734$, $p<.001$, $\eta_p^2=.296$). Post hoc analyses revealed that an expansion factor of 1.5 ($M=.797s$, $p=.015$) and 2.5 ($M=.753s$, $p=.024$) led to significantly longer movement time than an expansion factor of 1 ($M=.872s$); that is, no expansion effect applied. 

\begin{table}[tb]
  \caption{The results of RM-ANOVAs in the study.}
  \label{tab:RM-ANOVA}
  \scriptsize
	\centering%
  \begin{tabular}{ccccccc}
  \toprule
     & Factor & $DF_n$ & $DF_e$ & $F$ & $p$ & $\eta_p^2$ \\
    \midrule
    Movement & $ID$ & 2.466 & 39.460 & 3.776 & \textbf{.024} & .191\\
    Time & $E$ & 3 & 48 & 6.734 & \textbf{$<$.001} & .296 \\
         & $ID$$\times$$E$ & 3.200 & 51.203 & .709 & .560 & .042\\ \midrule
    Error & $ID$ & 2.712 & 43.396 & 1.238 & .306 & .072 \\
    Rate & $E$ & 3 & 48 & 2.613 & .062 & .140\\
         & $ID$$\times$$E$ & 15 & 240 & 1.045 & .409 & .061\\ \midrule
    Through- & $ID$ & 1.151 & 18.414 & 2.358 & .139 & .128\\
    put & $E$ & 1.186 & 18.979 & 2.105 & .162 & .116\\
        & $ID$$\times$$E$ & 1.261 & 20.169 & 1.206 & .299 & .070\\
  \bottomrule
  \end{tabular}
\end{table}

\begin{figure*}[htb]
 \centering 
 \includegraphics[width=\textwidth]{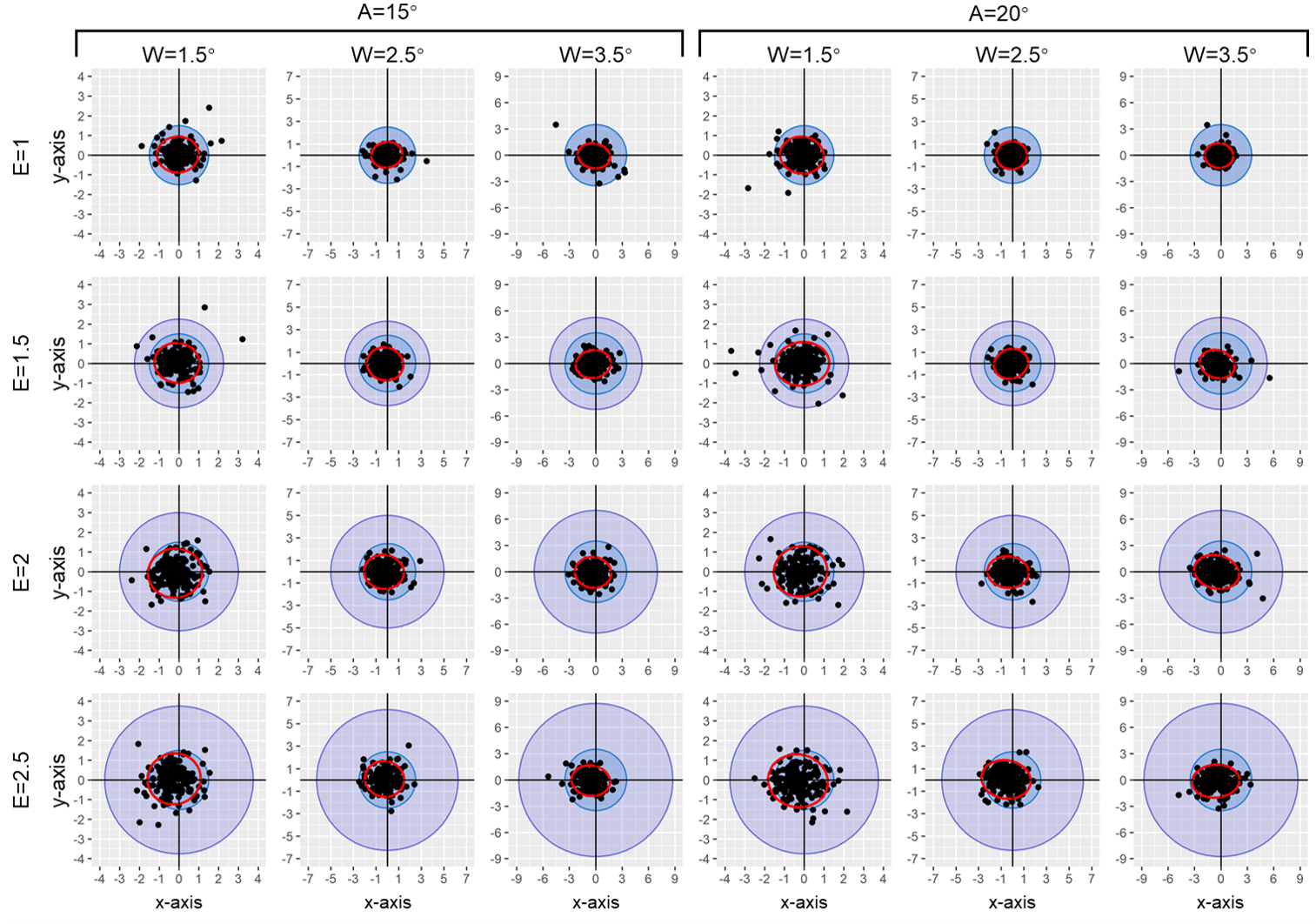}
 \caption{Distributions of endpoints across all participants in all conditions. $A$, $W$, and $E$ represent effective target width, effective movement amplitude, and expansion factor, respectively. The blue circles are the targets in initial size, the purple circles are the targets in expanded size, and the red ellipses are 95\% confidence ellipses. }
 \label{fig:endpoints}
\end{figure*}

\subsubsection{Fitts' Law Analysis and Endpoints}
Using Fitts' Law in \autoref{Eq:Fitts}, we found the MT (movement time) can be modeled as $MT=.186+.223*ID$ ($R^2=.972$) for $E=1$ condition, 
$MT=.206+.193*ID$ ($R^2=.963$) for $E=1.5$ condition, 
$MT=.112+.207*ID$ ($R^2=.984$) for $E=2$ condition, 
and $MT=.221+.173*ID$ ($R^2=.977$) for $E=2.5$ condition. 
\autoref{fig:barplotsandlm}(d) shows the Fitts' law results. 

% \begin{figure}[htb]
%  \centering 
%  \includegraphics[width=\columnwidth]{Figure-in-Use/LMPlot.png}
%  \caption{Fitts' law results for the four expansion factors. }
%  \label{fig:fitts}
% \end{figure}

We plotted all the endpoints in \autoref{fig:endpoints} to show their distributions in each condition.

\subsection{Answers to the Research Question}
According to the study results, expanding the target width by 1.5 and 2.5 times during the movement can significantly shorten the time for reaching the target. As expected, expanding the target size made the selection faster. However, no significant difference among expanding conditions was found (i.e., $E$ = 1.5, 2, and 2.5). One possible reason is that the expansion factors in our study did not vary largely from each other. Yet, for a large expansion factor, its application scenario would also be limited. Thus, we suggest VR developers consider expanding the target (e.g., buttons or 3D objects) to a suitable size to make the selection more efficient. 

We did not identify significant improvements in the accuracy and throughput for target selection when expanding targets. Notably, the main effect of $E$ on the error rate was close to statistically significant ($p=.062$, $\eta_p^2=.140$). From \autoref{fig:endpoints}, we can observe that expanding the target ($E$ = 1.5, 2, and 2.5) tolerated the selection pointed outside the initial target. While some endpoints were around the target in $E=1$ condition, which can be potentially selected if the target can expand. Future work with more participants, \textit{IDs}, and testing trials are needed to substantiate when the difference would appear. 

\section{Limitations and Future Work}
We only applied linear magnification effects to the target. Prior work has proposed non-linear magnification strategies~\cite{Gutwin2002Improving,Hornbaek2007Untangling}. Future work can also test them using Fitts' law approach to investigate the user motor performance. In addition, as a nature of the Fitts' law task, the target selection tasks were given repeatedly. It is inevitable that participants could predict the expansion of the target, which could affect their performance. We plan to conduct a follow-up study involving conditions in which the target could unpredictably expand, remain unchanged, or even shrink~\cite{Zhai2003Human} and test with more $IDs$ varying from a wider range~\cite{Soukoreff2004Towards}. Besides, prior work has been conducted to understand the endpoint distribution of pointing selection tasks~\cite{Yu2019Modeling, Wei2023Predicting}, which can help understand users' behavior from another perspective and can be integrated into selection techniques to improve the selection performance further. We intend to model endpoint distributions based on our collected data. We envision that with the support of predictive models, expanding the target can improve the selection performance in different VR scenarios (e.g., in dense scenarios where objects overlap~\cite{Shi2023Exploration} or outside of users' view~\cite{Yu2020Design}). 

\section{Conclusion}
In this paper, we conducted a Fitts' law study (\textit{N} = 17) to investigate how expanding the target affects the selection performance in virtual reality (VR) head-mounted displays (HMDs). Specifically, we tested three expansion factors (expand the target width to 1.5, 2, and 2.5 times its initial size) and compared them with a baseline non-expansion condition. Our results showed that expanding the target can improve the speed of target selection compared to the baseline condition. We then discussed the limitations of the present study, which could represent avenues for future work. This work serves as an initial exploration of expanding targets for pointing selection tasks in VR HMDs from a Fitts' law perspective and can help frame future work with relevant topics.

\acknowledgments{
The authors wish to thank the participants for their time and the reviewers whose comments helped improve our paper. This work was partly supported by the National Natural Science Foundation of China (\#62207022; \#62272396) and the Natural Science Foundation of the Jiangsu Higher Education Institutions of China (\#22KJB520038).}

\bibliographystyle{abbrv-doi}
%\bibliographystyle{abbrv-doi-narrow}
%\bibliographystyle{abbrv-doi-hyperref}
% \bibliographystyle{abbrv-doi-hyperref-narrow}

% \item For adding hyperlinks and DOIs to the list of references, you can use ``\texttt{\textbackslash bibliographystyle\{abbrv-doi-hyperref-narrow\}}'' (instead of ``\texttt{\textbackslash bibliographystyle\{abbrv\}}''). It uses the doi and url fields in a bib\TeX\ entry and turns the entire reference into a link, giving priority to the doi. The doi can be entered with or without the ``\texttt{http://dx.doi.org/}'' url part. See the examples in the bib\TeX\ file and the bibliography at the end of this template.

\bibliography{Reference-in-Use}
\end{document}